\documentclass[doublecol]{epl2} 

\usepackage{tabularx}
\usepackage{amsmath}
\usepackage{amsfonts}
\usepackage{graphicx}

\newcommand{\average}[1]{\ensuremath{\langle#1\rangle} }

\title{Evolutionary prisoner's dilemma games on the network with punishment and opportunistic partner switching}
\shorttitle{Prisoner's dilemma on the network with punishment and partner switching} 

\author{H. Takesue \inst{1}}
\shortauthor{H. Takesue}

\institute{ 
\inst{1} Graduate Schools for Law and Politics, The University of Tokyo - Hongo, Bunkyo, Tokyo 113-0033, Japan\\
}

\abstract{
Punishment and partner switching are two well-studied mechanisms that support the evolution of cooperation. Observation of human behaviour suggests that the extent to which punishment is adopted depends on the usage of alternative mechanisms, including partner switching. In this study, we investigate the combined effect of punishment and partner switching in evolutionary prisoner's dilemma games conducted on a network. In the model, agents are located on the network and participate in the prisoner's dilemma games with punishment. In addition, they can opportunistically switch interaction partners to improve their payoff. Our Monte Carlo simulation showed that a large frequency of punishers is required to suppress defectors when the frequency of partner switching is low. In contrast, cooperation is the most abundant strategy when the frequency of partner switching is high regardless of the strength of punishment. Interestingly, cooperators become abundant not because they avoid the cost of inflicting punishment and earn a larger average payoff per game but rather because they have more numerous opportunities to be referred as a role agent by defectors. Our results imply that the fluidity of social relationships has a profound effect on the adopted strategy in maintaining cooperation.}

\begin{document}

\maketitle

\section{Introduction}
The evolution of cooperation is an actively studied subject in the biological and physical sciences \cite{Nowak2006, Szabo2007a, Perc2017}. Social efficiency can be enhanced if each individual pays the cost of cooperation in helping other individuals. However, individuals can maximise their own benefit by enjoying the fruits of others' cooperative behaviour without paying the cost of cooperation. This temptation to free-ride on the cooperation of others makes cooperation vulnerable against defectors. The tension between social efficiency and individual welfare is called `social dilemma' and is analysed within the framework of (evolutionary) game theory. 

Many theoretical models have been proposed to understand ways in which the discrepancy between individual and group benefits is resolved and to explain the evolutionary origin of cooperation. Proposed mechanisms include direct reciprocity \cite{Axelrod1981}, indirect reciprocity \cite{Nowak2005a}, voluntary participation \cite{Hauert2002}, tag-based cooperation \cite{Fu2012} and network structure \cite{Nowak1992, Santos2005}, among several others. 

In this paper, we examined the combined effect of two mechanisms adopted in the previous literature--punishment and partner switching. Punishment is an additional strategy that is appended to standard social dilemma games, the strategies of which are limited to cooperation or defection (free-riding). Punishers not only pay the cost of cooperation but also incur a cost in punishing defectors. Experimental studies have shown that appending a punishment strategy enhances the level of cooperation \cite{Fehr2002}. This observation raises another evolutionary puzzle because there are no incentives for each individual to bear the cost of punishment. Cooperators in particular are called second-order free-riders because they only incur the cost of cooperation and avoid the burden of imposing punishment \cite{Oliver1980}. Researchers have explained the evolution of punishment by referring to group selection \cite{Boyd2003}, reputation \cite{Hilbe2012, DosSantos2011a}, and spatial interaction \cite{Helbing2010, Helbing2010c, Helbing2010d, Szolnoki2011b, Wang2013c, Chen2014, Chen2015, Szolnoki2017a}. 

Despite its importance, punishment is not always the dominant solution in solving the social dilemma problem when other options are available \cite{Ohtsuki2009}. Field research of human behaviour has indicated that the severing of social ties with free-riders is more widely observed than the costly imposition of punishment \cite{Guala2012}. The effects of the modifications of social relationships have been studied extensively under the framework of the co-evolutionary game \cite{Perc2010}. Models of the evolutionary games on the (social) network assume that agents are located on the network and that game interactions occur with neighbours who are connected by links \cite{Nowak1992, Santos2005}. Co-evolutionary game models consider the possibility that agents sever the links with current neighbours and construct new links on the basis of the strategy and payoff from the games in which they have participated \cite{Zimmermann2004, Eguiluz2005, Pacheco2006a, Santos2006, Fu2007, Fu2008, VanSegbroeck2008, Fu2009, Szolnoki2009b, Wu2009c, Zhang2011, Li2013a, Yang2013a, Cong2014, Li2014a, Xu2014, Chen2016, Pinheiro2016, Wang2016f, Li2017b}. Many of these models showed that sufficient opportunities for link adaptation can foster the evolution of cooperation. Human experiments also showed that opportunities for partner switching induce higher cooperation levels \cite{Rand2011, Gallo2015, Cuesta2015}. The co-evolutionary framework is applied to other social phenomena, including consensus formation \cite{Fu2008a} and the evolution of fairness \cite{Deng2011, Gao2011, Takesue2017a}. 

Here we introduce the partner switching mechanism and punishment strategy to the prisoner's dilemma game (PDG) on a network. By conducting Monte Carlo experiments we showed that punishers can repress defectors without partner switching if the strength of the punishment is high. In contrast, cooperation is the most abundant strategy when the frequency of link adaptation is high regardless of the strength of the punishment. Notably, a high frequency of cooperators is achieved without their gaining a higher average payoff per game than that gained by punishers. Our simulation outcomes--to be shown in detail shortly--imply that the distinction between punishers and second-order free-riders is blurred in the co-evolutionary PDG. 

In what follows, we first explain the PDG with network evolution and punishment. Then, we present the results of the simulation. In the last section, we discuss the implications of our results. 

\section{Model}
In the standard version of the PDG, agents have two strategies: cooperation ($C$) and defection ($D$). Cooperators incur a cost in helping other agents, while defectors do not. The burden of the cooperation cost leads to the cooperators being at a payoff disadvantage in comparison with defectors, and cooperators become extinct without the support of additional mechanisms. In this study, we use the extended version of the PDG that includes the punishment ($P$) strategy. Punishers pay the cost of cooperation in the same manner that cooperators do. In addition, after observing the partner's choice ($C$ or $D$), punishers incur the cost of sanction ($\gamma$) to apply a fine ($\alpha$) to defectors. Using the rescaled payoff matrix \cite{Nowak1992} and appending the punishment strategy, the payoff matrix of a studied game is given by \cite{Wang2013c}:
\begin{eqnarray}
\label{payoff_matrix}
\bordermatrix{
& C & D & P\cr
C & 1 & 0 & 1 \cr
D & b & 0 & b-\alpha \cr
P & 1 & -\gamma & 1 \cr
},
\end{eqnarray}
where $b$ is the temptation to defect ($1 < b < 2$). Cooperators are called second-order free-riders because they do not incur the cost of punishment. 

Next, we explain the evolutionary process of the strategy and network structure within which the game is played. We consider a set of $N$ agents who are placed on the network. Links between nodes (agents) represent social relationships (game interactions) existing between two agents. Initially, agents are located on a homogeneous network (average degree \average{k}, degree heterogeneity appears after link adaptation), where each agent's neighbour is randomly decided \cite{Santos2005a}. Each agent's strategy is uniformly distributed. 

The Monte Carlo simulation consists of two kinds of events: strategy updating events and partner switching events. Strategy updating events occur with the probability $1-w$. In strategy updating events, one link is chosen randomly \cite{Fu2009}, and the roles of two connected agents (focal or role) are also determined randomly. The focal agent may imitate the role agent's strategy on the basis of the comparison of payoffs from games. In particular, each agent participates in PDG with punishment with all the neighbours and accumulates payoffs. Next, the payoffs are compared and the focal agent imitates the strategy of the role agent with a probability derived by the following equation:
\begin{equation*}
P(s_f \rightarrow s_r) = [1 + \exp(-\beta(\Pi_r - \Pi_f))]^{-1},
\end{equation*}
where $\Pi_f$ ($\Pi_r$) denotes the accumulated payoff of the focal (role) agent. The $\beta$ is the intensity of selection ($\beta \to 0$ implies the random adoption of a strategy, whereas $\beta \to \infty$ implies the deterministic imitation). In the following analysis, we set the value of $\beta$ to 0.1, which implies that agents tend to adopt a strategy that results in a higher payoff, but the possibility of `irrational' imitation is also allowed. We assume that with probability $\mu$, mutation occurs in strategy updating events and that a randomly selected agent from the population adopts one strategy uniformly \cite{Helbing2010d}. The possibility of mutation ensures that the resultant distribution of strategies is robust against small perturbations. 

Partner switching (link adaptation) events occur with the probability $w$, and one randomly chosen focal agent may cut a relationship with one randomly selected current neighbour and construct a relationship with one randomly selected non-neighbour. Following standard literature \cite{Perc2010}, we assume that agents modify social relationships on the basis of the strategy that other agents adopt. In particular, the focal agent severs the link with one randomly selected current neighbour and reconnects that link with one randomly selected non-neighbour (potential new neighbour) if the rewiring realises the strict improvement in payoff (combinations of strategies that result in link adaptation are shown in Table \ref{tab_link}; see also payoff matrix (\ref{payoff_matrix})). In this regard, agents depend on local knowledge (information on neighbours' strategies) as well as reputational knowledge (information on non-neighbours' strategies) \cite{Gallo2015}. 
\renewcommand{\arraystretch}{1.15}
\begin{table}[tbp]
\caption{\small
The combinations of the strategy of a focal agent, a randomly selected current neighbour and a randomly selected potential new neighbour (current non-neighbour) that lead to link adaptation are shown. Link adaptation does not occur in other cases because payoff improvement is not achieved.}
\centering
\begin{tabularx}{\linewidth}{p{5em}p{4.5em}p{11em}}
\hline
\hline
\multicolumn{3}{c}{The strategy of a} \\
\hfil focal agent & \hfil neighbour & \hfil potential new neighbour \\[0.1ex]
\hline
\hfil $C$ & \hfil $D$ & \hfil $C$, $P$ \\ 
\hfil $D$ & \hfil $D$ & \hfil $C$, $P$ (if $b - \alpha > 0)$ \\
&& \hfil \ \ $C$ \ \ (if $b - \alpha < 0$) \\
\hfil $D$ & \hfil $P$ & \hfil \ \ $C$ \ \ (if $b - \alpha > 0$) \\
&& \hfil $C$, $D$ (if $b - \alpha < 0)$ \\
\hfil $P$ & \hfil $D$ & \hfil $C$, $P$ \\
\hline
\hline
\end{tabularx}
\label{tab_link}
\end{table}
\renewcommand{\arraystretch}{1}

In link adaptation events, for example, a focal cooperator (punisher) switches partners when a selected neighbour is a defector and a selected non-neighbour is a cooperator or a punisher because an interaction with this player (a new neighbour) results in a larger payoff. This is the same for a focal defector if we assume that the fine of punishment is not large enough for a focal defector to prefer defectors over punishers in game interactions ($b-\alpha > 0$). In addition, a defector can sever the partnership with a punisher and establish a link with a cooperator. In contrast, when the fine of punishment is large ($b-\alpha < 0$), defectors prefer cooperators over defectors and defectors over punishers and modify partnerships accordingly. Partner switching does not occur in other cases because the link rewiring lowers the payoff or does not change the situation. We call this link adaptation rule `opportunistic partner switching' because agents seek to improve their gain by finding more beneficial interactions. We report mainly the results with $b-\alpha > 0$ because the basic result is the same in either case. We briefly mention the case where $b-\alpha < 0$ at the end of the Results section and explain why both cases reach similar results. In the simulation, we imposed the restriction that the link to an agent who has only one neighbour may not be severed so that all agents can participate in the PDG \cite{Santos2006}.

\section{Results}
To investigate the results of the evolutionary process, we conducted Monte Carlo simulations. The simulation continued for $10000 \times N$ periods in the first place, and we averaged the values of following $20000 \times N$ periods to calculate the quantities of interest. Our main interest is in the frequency of each strategy ($\rho_C$, $\rho_D$ and $\rho_P$). We report the average of five independent simulations for each combination of parameters to enhance statistical accuracy. Because the model includes many parameters, supplementary material reports the results with different values of parameters and checks the robustness of the patterns explained below.

\begin{figure}[tbp]
\centering
\vspace{-5mm}
\includegraphics[width = 70mm, trim= 0 20 0 -20]{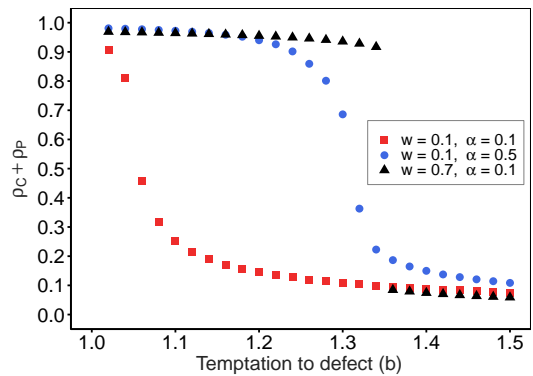}
\caption{\small Sum of the proportion of cooperators and punishers ($\rho_C + \rho_P$) as a function of the temptation to defect ($b$). Both harsher punishment and higher frequency of partner switching help cooperators and punishers. Parameters: $N = 50 000, \average{k} = 8, \gamma = 0.05, \beta = 0.1, \mu = 0.005$.}
\label{rho_b_wa}
\end{figure}

First, in providing an overview of the basic results of the simulation, we report the frequency of cooperative individuals ($\rho_C + \rho_P$) as a function of the temptation to defect ($b$). Figure~\ref{rho_b_wa} shows that cooperation deteriorates with small $b$ when partner switching is slow ($w = 0.1$) and the punishment fine is small ($\alpha = 0.1$). Defectors who exploit cooperators and punishers to achieve a larger payoff can proliferate in the population. The figure also shows that both partner switching and harsh punishment can help cooperative agents. A more gradual decrease of cooperators and punishers is observed with harsher punishment ($\alpha = 0.5$), and cooperative agents flourish with larger values of $b$. Faster partner switching ($w = 0.7$) also helps cooperators and punishers: they predominate over defectors with larger $b$ until a discontinuous drop was observed. 

\begin{figure}[tbp]
\centering
\vspace{-5mm}
\includegraphics[width = 80mm, trim= 0 10 0 0]{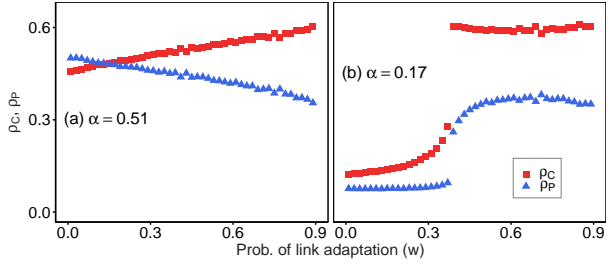}
\caption{\small Proportion of cooperators and punishers ($\rho_C$ and $\rho_P$) as a function of the frequency of partner switching ($w$). Panel (a) shows that only strong punishment is sufficient to suppress defectors. Panel (b) shows that fast link adaptation can also help cooperators and punishers even with weak punishment. Cooperators increases in frequency when $w$ is large regardless of the value of $\alpha$. Parameters: (a) $\alpha = 0.51$, (b) $\alpha = 0.17$; fixed $N = 50 000, \average{k} = 8, b = 1.15, \gamma = 0.05, \beta = 0.1, \mu = 0.005$.}
\label{rho_w_a}
\end{figure}

To elaborate the effects of partner switching and punishment, we report $\rho_C$ and $\rho_P$ as a function of $w$ for different values of $\alpha$ in fig.~\ref{rho_w_a}. Panel (a) shows the result of harsh punishment ($\alpha = 0.51$), and only the existence of a strong punisher is sufficient to outperform defectors. Defectors are suppressed without fast partner switching (small $w$). In this situation, a certain number of punishers is required because cooperators cannot resist the invasion by mutant defectors. Defectors acquire higher payoff in each game and cooperators diminish their frequency without the support of punishers. The panel also shows that the frequency of cooperators increases as the value of $w$ becomes larger. Panel (b) shows the results of weak punishment ($\alpha = 0.17$), and defectors proliferate when $w$ is small. However, a discontinuous increase in $\rho_C$ and $\rho_P$ is observed as $w$ exceeds the specific value. In addition, the cooperator is the winner of the evolutionary process in this phase, and the frequency of cooperators is larger than that of punishers. The frequency of cooperators is about twice that of punishers with this combination of parameters.

\begin{figure}[tbp]
\centering
\vspace{-5mm}
\includegraphics[width = 80mm, trim= 0 20 0 7]{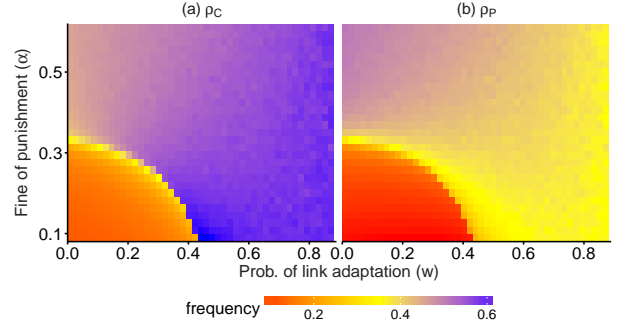}
\caption{\small Panel (a) shows the proportion of cooperators ($\rho_C$) and Panel (b) shows the proportion of punishers ($\rho_P$) as a function of ($w, \alpha$). Strong punishment is required to suppress defectors when $w$ is small. Cooperation is the most abundant strategy when $w$ is large. Parameters: $N = 50 000, \average{k} = 8, b = 1.15, \gamma = 0.05, \beta = 0.1, \mu = 0.005$.}
\label{phase}
\end{figure}

To provide the comprehensive explanation of the patterns that were observed in fig.~\ref{rho_w_a}, we report $\rho_C$ and $\rho_P$ in the $w-\alpha$ space in fig.~\ref{phase}. When $w$ is small evolutionary outcomes are dependent on the strength of punishment: larger $\alpha$ is required for the evolution of cooperation and punishment. Note that the required efficiency of punishment ($\alpha/\gamma$) depends on the cost of punishment ($\gamma$). In fig.~\ref{phase}, relatively efficient punishment ($\gamma = 0.05$ and $\alpha \approx 0.35$) is required for the evolution of cooperators and punishers. In contrast, we confirmed that, when $\gamma$ is large ($\gamma = 0.30$), relatively less efficient punishment ($\alpha = 0.88$) is sufficient to repress defectors without partner switching.

When $w$ is sufficiently large, defectors diminish in frequency regardless of the size of $\alpha$. This result is not surprising because many previous studies have shown that fast partner switching can support the evolution of cooperation without punishment. The novel finding here is that the winners of the evolutionary process in this situation are cooperators. 

\begin{figure}[tbp]
\centering
\vspace{-5mm}
\includegraphics[width = 70mm, trim= 0 20 0 0]{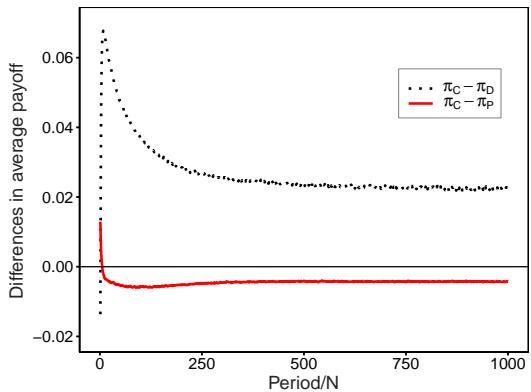}
\caption{\small Payoff difference ($\pi_C-\pi_D$ and $\pi_C-\pi_P$) as a function of time. We calculated the mean values for every $N$ period. Payoff for cooperators exceeds that of defectors in the early rounds of the simulation. In contrast, the advantages for cooperators over punishers disappear at the early stages of the simulation. Figure \ref{payoff} and \ref{migrate} report the average of 200 simulation runs. Parameters: $N = 50 000, \average{k} = 8, b = 1.15, \gamma = 0.05, \alpha = 0.17, w = 0.4, \beta = 0.1, \mu = 0.005$.}
\label{payoff}
\end{figure}

The result indicating that cooperators outperform punishers when the frequency of partner switching is high may seem to be a natural one because cooperators are second-order free-riders, and they do not incur the cost of punishment. However, the mechanism that supports the proliferation of cooperation is more intricate. To understand the evolutionary process, we first compute the average payoff per game for each strategy ($\pi_C$, $\pi_D$, $\pi_P$) and report the time-series of their differences ($\pi_C - \pi_D$ and $\pi_C - \pi_P$) in fig.~\ref{payoff}. The figure shows that cooperators earn a larger payoff than do defectors. Although defectors earn a larger payoff by free-riding at the outset of the simulation, this advantage deteriorates rapidly. Previous studies have pointed out that an assortment of cooperators is the key to the success of the cooperators in evolutionary games on the network and that the assortment is helped by partner switching because cooperators can sever the relationships with defectors \cite{Fu2008}. 

The figure also shows that cooperators earn a larger payoff than punishers during the early stages of the simulation process. Because only punishers incur the cost of punishment, second-order free-riders (that is, cooperators) acquire larger payoff. However, the advantage for cooperators disappears quickly, and they earn a \textit{smaller} payoff than punishers. (This situation does not appear in the case of well-mixed populations where punishers gain a smaller average payoff than do cooperators as long as defectors exist.)

This disadvantage for cooperators is the result of the opportunistic link adaptation mechanism. To achieve a larger payoff by avoiding the fine of punishment, defectors sever their relationships with punishers and create new links with cooperators. Consequently, punishers can avoid the burden of interactions with defectors although they earn a smaller payoff than cooperators \textit{once} they participate in interactions with defectors. The co-evolutionary process thus improves the total payoff as well as average payoff per game of punishers. In addition, although cooperators do not suffer from decreases in total payoff due to partnerships with defectors, link adaptation leads to the transfer of the risk of invasion by defectors to cooperators. The partner switching mechanism thus changes the situations where cooperators just free-ride on punishers. 

\begin{figure}[tbp]
\centering
\vspace{-5mm}
\includegraphics[width = 75mm, trim= 0 20 0 0]{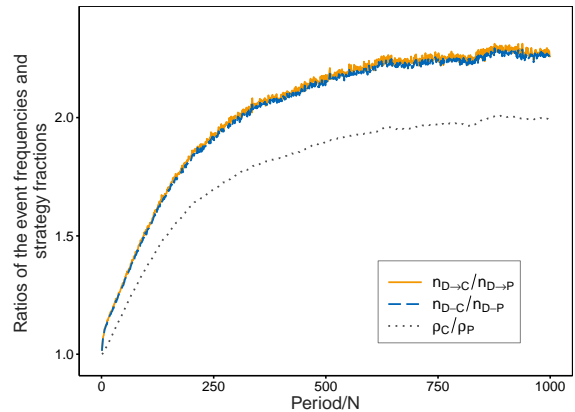}
\caption{\small The ratio of the number of times defectors adopt cooperation ($n_{D\to C}$) to the number of times defectors adopt punishment ($n_{D\to P}$) as a function of time. We calculated the mean values for every $N$ period. Cooperators are more likely to enforce their strategy on punishers. Parameters: $N = 50000, \average{k} = 8, b = 1.15, \gamma = 0.05, \alpha = 0.17, w = 0.4, \beta = 0.1, \mu = 0.005$.}
\label{migrate}
\end{figure}

Paradoxically, links with defectors support the evolutionary success of cooperators. Because defectors' relationships with cooperators and punishers are severed, they suffer from a lower average payoff per game, as well as a smaller number of games in which they can participate (as we see below, only fast partner switching is sufficient to suppress the proliferation of defectors). Consequently, the combination of link adaptation and punishment (or very fast link adaptation alone) helps cooperators and punishers to earn a larger payoff than defectors and facilitates transitions from defection to cooperation or punishment. However, cooperators enjoy the larger benefit from this transition. We counted the number of times defectors adopt cooperation ($n_{D\to C}$) and punishment ($n_{D\to P}$) and report the time-series of their ratio ($n_{D\to C}/n_{D\to P}$) in fig.~\ref{migrate}. The figure shows that the number of transitions from defectors to cooperators is larger than that from defectors to punishers. The advantage for cooperators is prominent even if we consider the ratio of strategy frequency ($\rho_C/\rho_P$) that is also plotted in fig.~\ref{migrate}. Although more cooperators than punishers adopt a defective strategy, the larger number of relationships with defectors leads to the proliferation of cooperators because the absolute number of transitions to cooperator is larger than that of transitions to defector. 

This advantage for cooperators stems from the fact that partnership with defectors supports cooperators by securing larger opportunities to enforce their strategy on defectors. Because defectors sever their relationships with punishers, defectors and punishers tend to be segregated and strategy transmission between these two types of agents becomes rare. In contrast, since those with severed links to punishers are connected to cooperators, cooperators have a greater number of opportunities to be referred to as a role agent and for their strategies to be copied by defectors. These patterns are verified by the ratio of $n_{D-C}$ to $n_{D-P}$. Here, $n_{D-C}$ ($n_{D-P}$) denotes the number of times that strategy updating events between a focal defector and a role cooperator (punisher) occured. As shown in fig.~\ref{migrate}, strategy updating events between focal defectors and role cooperators occur more frequently and cooperators have larger opportunities to enforce their strategy. Once the payoff dominance over defectors is achieved, a link with defectors brings cooperators the opportunities of proliferation rather than the risk of invasion. 

In addition, connection with defectors is beneficial for cooperators regarding the number of games in which they can participate. In the short term, interactions with defectors result in zero payoff and are not immediately beneficial for cooperators. However, once defectors turn into cooperators, cooperators can enjoy the positive payoff from a larger number of games, which also helps cooperators to effectively enforce their strategy on defectors. Figure~\ref{migrate} shows that $n_{D\to C}/n_{D\to P}$ is slightly larger than $n_{D-C}/n_{D-P}$, which implies that cooperators who earn a larger payoff exploit the opportunities of strategy enforcement more effectively. 

We confirmed that the same pattern is observed when $\alpha$ is large although the result is not reported here. Because harsh punishment diminishes the number of defectors, the effective frequency of link adaptation is lowered. However, the advantage of being imitated by defectors more frequently is achieved when $w$ is large enough and cooperators proliferate as we observed in figs.~\ref{rho_w_a} and \ref{phase}.

\begin{figure}[tbp]
\centering
\vspace{-5mm}
\includegraphics[width = 80mm, trim= 0 20 0 -5]{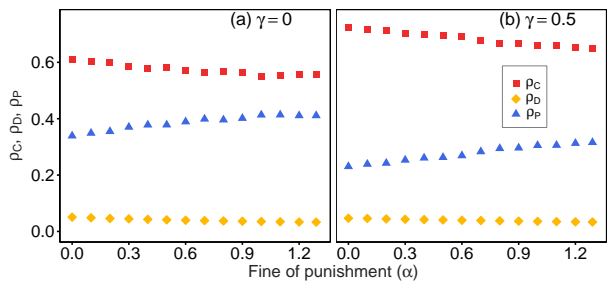}
\caption{\small Proportions of cooperators and punishers ($\rho_C$ and $\rho_P$) as a function of the fine of punishment ($\alpha$) for the different cost of punishment ($\gamma$). With fast link adaptation, cooperators achieve a larger frequency than do punishers without the cost of punishment, and this difference becomes larger with the cost becoming positive. Parameters: $N = 50 000, \average{k} = 8, b = 1.3, w = 0.8, \beta = 0.1, \mu = 0.005$.}
\label{rho_a_g}
\end{figure}

To confirm the robustness of the result that fast link adaptation supports the dominance of cooperators, we investigate cases where the cost of punishment ($\gamma$) is small. Note that for the enhancement of comparability, we assume that partner switching is implemented in the same manner (i.e. link adaptation occurs as if $\alpha > 0$) although we often examine cases where $\alpha = 0$ and defectors gain same payoff from interactions with cooperators and punishers. Figure~\ref{rho_a_g} reports the proportion of each strategy as a function of $\alpha$ for different values of $\gamma$. First, the figure shows that fast partner switching can diminish the frequency of defectors without substantive punishment ($\alpha = 0$), which implies that only large $w$ excludes the defection of candidates from winning strategies. Second, panel (a) shows that frequency of cooperators is larger than that of punishers when $\gamma = 0$. This result corroborates the interpretation that the cost of implementing punishment is not the main reason for the predominance of cooperators. Third, larger $\alpha$ contributes to the gradual decrease in the number of defectors and mitigates the disadvantage of punishers relative to cooperators because defectors are more likely to become cooperators with fast link adaptation. However, this tendency does not reach the reversal of $\rho_C$ and $\rho_P$. Last, panel (b) shows that the large cost of implementing punishment ($\gamma = 0.5$) amplifies the abundance of cooperators as expected.

\begin{figure}[tbp]
\centering
\vspace{-5mm}
\includegraphics[width = 80mm, trim= 0 20 0 -10]{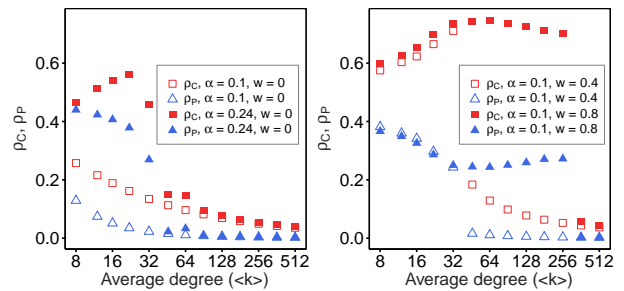}
\caption{\small Proportions of cooperators and punishers ($\rho_C$ and $\rho_P$) as a function of the average degree ($\average{k}$). Cooperators and punishers decrease their frequency simultaneously with larger $\average{k}$. Average of 100 simulations is reported when $(\average{k}, \alpha, w) = (32, 0.24, 0)$. Parameters: $N = 50 000, b = 1.05, \gamma = 0.05, \beta = 0.1, \mu = 0.005$.}
\label{rho_k}
\end{figure}

The analysis has so far assumed that the network has low density. Here we examine different cases by varying the average degree (\average{k}). Panel (a) of fig.~\ref{rho_k} reports the results without link adaptation. In the typical standard prisoner's dilemma games on the network, a large degree makes it difficult to form the cluster of cooperators and deteriorates cooperation. The same pattern is observed when the punishment strategy is appended and a large average degree deteriorates the cooperators and punishers. Panel (b) reports the results with link adaptation. The pattern that cooperators achieve a larger frequency than do punishers is confirmed again when $\average{k}$ is small. However, a larger $\average{k}$ supports defectors, and cooperators and punishers decrease their frequency simultaneously.

\begin{figure}[tbp]
\centering
\vspace{-5mm}
\includegraphics[width = 65mm, trim= 0 20 0 -10]{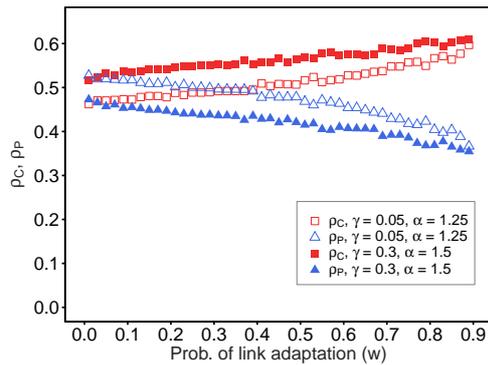}
\caption{\small Proportions of cooperators and punishers ($\rho_C$ and $\rho_P$) as a function of the frequency of partner switching ($w$). The pattern in Panel (a) of fig.~\ref{rho_w_a} is replicated. Parameters: $N = 50 000, \average{k} = 8, b = 1.15, \beta = 0.1, \mu = 0.005$.}
\label{large_fine}
\end{figure}

Last, we consider the case where the fine of punishment is larger than the temptation to defect ($b - \alpha < 0$) and defectors prefer interactions with defectors over those with punishers. Figure~\ref{large_fine} reports the resultant frequency of each strategy as a function of $w$ when punishment is strong. The pattern that was observed in Panel (a) of fig.~\ref{rho_w_a} is replicated: punishers are required when $w$ is small, whereas cooperators achieve a larger frequency when $w$ is large. Although focal defectors prefer interactions with defectors over those with punishers, cooperators remain the most preferred interaction partner. Consequently, cooperators still have a large number of opportunities to enforce their strategy on defectors when $w$ is large. 

\section{Discussion}
In this paper, we examined the evolutionary PDG with punishment and partner switching. Our simulation results showed that harsh punishment can suppress the proliferation of defectors even when the frequency of partner switching is low. Defectors also diminish in frequency when the frequency of partner switching is high, and cooperators outperform punishers in this situation. The advantage of cooperators in this situation is not derived from the fact that they do not incur the cost of punishment and acquire larger average payoff per game. In well-mixed populations or on the static network, cooperators are just second-order free-riders. With opportunities for partner switching, however, defectors try to sever their relationship with punishers and make connections with cooperators. This network evolution diminishes cooperators' payoff advantage over punishers. The advantage for cooperators results from the fact that they have more opportunities to enforce their strategy on defectors, thereby increasing the frequency of cooperators in their neighbourhood. 

Our results showed that the abundance of each strategy in social dilemma differs depending on the frequency of partner switching. This fact suggests that the fluidity of social relationships has a profound effect on what strategy is favoured in the evolutionary process. A simple cooperation strategy is favoured with a high level of social fluidity, whereas the punishment strategy is favoured in the society with fixed social relationships. In addition, our model suggests that the normative distinction between punishers and second-order free-riders is not so clear because punishers can avoid interactions with defectors. This observation might be related to the fact that punishers do not necessarily enjoy positive reputations \cite{Raihani2015}. 

Although our model builds on the understanding of the combined effect of partner switching and punishment, further analysis is warranted. For example, anti-social punishment is indicated as another obstacle to the evolution of cooperation (punishment) \cite{Herrmann2008} and an investigation into its effects may be required. In addition, adding a reward mechanism \cite{Chen2014a} may lead to an interesting pattern. Furthermore, since the number of opportunities to enforce the strategy played a large role in the model, introducing other rules of strategy updating may be required \cite{Cardillo2010, Cimini2014, Szolnoki2015}. We believe that this line of research will further deepen our understanding of the evolutionary origin of cooperation.


\end{document}